\documentclass[a4paper]{jpconf}
\usepackage{graphicx}
\usepackage{amsmath, bm}
\begin{document}
\title{The remains of a spinning, hyperbolic encounter}

\author{L De Vittori$^1$, A Gopakumar$^2$, A Gupta$^2$, P Jetzer$^1$}

\address{$^1$ Physik-Institut, Universit\"at Z\"urich, 8057 Z\"urich, Switzerland\\$^2$ Department of Astrophysics, Tata Institute of Fundamental Research, Mumbai, India}

\begin{abstract}
  We review a recently proposed approach to construct gravitational wave (GW) polarization states of unbound spinning compact binaries.
 Through this rather simple method, we are able to include corrections due to the dominant order spin-orbit interactions, in the quadrupolar approximation and in a semi-analytic way.
 We invoke the 1.5 post-Newtonian (PN) accurate quasi-Keplerian parametrization for the radial part of the dynamics and impose its temporal evolution in the PN accurate polarization states equations.
 Further, we compute 1PN accurate amplitude corrections for the polarization states of non-spinning compact binaries on hyperbolic orbits.
 As an interesting application, we perform comparisons with previously available results for both the GW signals in the case of non-spinning binaries and the theoretical prediction for the amplitude of the memory effect on the metric after the hyperbolic passage.
\end{abstract}

\section{Introduction}
In preparation of the first direct observations of GWs with ground-based detectors, expected in the next few years,
it is desirable to have GW templates for every possible initial configuration of the two-body problem, and that as precise as possible.
Our recent work belongs to that program, investigating in particular the gravitational radiation from hyperbolic encounters and its effects on the metric.

Previous work on the subject includes the Newtonian treatment of the gravitational radiation
field associated with non-spinning objects on unbound orbits, done by Turner \cite{T77},
it's first extension to the first PN order in the dynamics done by Junker \& Sch\"afer \cite{JS92},
the full 1PN corrected waveforms, including PN accurate polarization states $h_+$ and $h_{\times}$,
available in \cite{MFV10}, where a generalized true anomaly parametrization was employed.

No treatment of spinning binaries on hyperbolic orbits can be found in the literature,
a part from the work we review here, by De Vittori et al. \cite{DGGJ14}.
More details about GW radiation were described for the non-spinning case only,
e.g. the quadrupolar order energy and momentum loss during the interaction \cite{WW76}
and its 1PN extension \cite{BS89}, or the analytic formula for the power spectrum \cite{DKJ12},
which generalizes the high eccentricity limit \cite{T77} and the parabolic limit \cite{BG10}.

In this summary we review our recent work \cite{DGGJ14}, where we obtained temporally evolving GW
polarization states for spinning compact binaries in PN accurate hyperbolic orbits.
We included leading order spin-orbit interactions effects, and the conservative
non-spinning orbital dynamics is 1PN accurate.
Therefore, we were able to describe spinning compact binaries along unbound orbits
up to 1.5 PN order accuracy. Additionally, we incorporated 2.5 PN radiation reaction effects
while computing the quadrupolar order polarization states $h_+$ and $h_{\times}$.

\subsection{The Memory Effect}
The plots for GW signals at PN order reveal a non-vanishing difference between the wave
amplitudes at $t=\infty$ and $t=-\infty$, which is called the linear \emph{memory effect}.
In the following, we will review the basic concepts of this memory, as well as some
crude estimates of its amplitude at Newtonian order, available in Refs.~\cite{ZP74, BG85, BT87}
and briefly summarized in \cite{F09}.

Consider an ideal GW detector in which truly free-falling test masses are
only sensitive to gravitational forces, like e.g. a ring of test masses in space,
a Pulsar Timing Array (PTA), or a LISA-like detector.
A GW without memory causes the detector to return to its initial configuration.
Instead, after the passage of a GW with memory the displacement state of the
detector is different than long before the wave had passed.
This net change in the amplitude of the metric induced by a GW
is what we define as the memory effect.

Ground based detectors are not ideal detectors and this kind of memory
cannot be stored in their signal, since internal forces, e.g. the wires on the pendulum,
bring the test masses back to their original positions.
However, for new generation detectors and especially for PTA this signature should be taken
seriously into account since it appears in the polarization states already at Newtonian order.

The quadrupolar order GW polarization states we require for a system of
two orbiting masses $m_1$ and $m_2$ at a distance $R$, can be expressed through the
transverse-traceless part of the radiation field
$h_{ij}^\text{TT}$, where the quadrupolar order contribution, as is well-known, reads
\begin{equation}\label{eq:definition_h_newton}
  h_{ij}^\text{TT} \big|_{\text Q} = \frac{2\, G}{ c^4 R}\, \ddot{\mathcal{I}}_{ij}^{TT}~,
\end{equation}
$G$ being the gravitational constant, $c$ the speed of light, and where
$\mathcal{I}_{ij}^{TT} = \mu (x_i x_j)^{TT}$ is the source mass-quadrupole moment,
with the reduced mass $\mu=m_1\,m_2/m$, the total mass $m$ and the relative orbital separation
components $x_i$.
Using the equation of motion $\ddot{x}_i= -G m x_i/r^3$,
the second derivative of the mass-quadrupole moment reads
\begin{equation}\label{eq:Iij_derivative}
  \ddot{\mathcal{I}}_{ij} = 2 \mu (\dot{x}_i\,\dot{x}_j - \frac{GM}{r^3}x_i\,x_j)~,
\end{equation}
and one can notice that the second term in eq.~(\ref{eq:Iij_derivative})
vanishes for $t\to\pm\infty$ since it falls off as $1/r$, while the relative
velocities approach the finite value ${\rm v}_{\infty}=\sqrt{2E/\mu}$.

Therefore, at Newtonian order the memory effect for a hyperbolic encounter
can be predicted through the simple formula
\begin{equation}\label{eq:general_memory}
  \Delta h_{ij} = 4\frac{G \mu}{c^4 R} \Delta(\dot{x}_i\,\dot{x}_j)~,
\end{equation}
where $\Delta$ denotes the difference between the initial and final state.
The velocity directions at early and late times will thus obviously play a key role.

Let's suppose that the observer is perpendicular to the orbital plane,
say on the $x_3$ axis. Then only the components $\mathcal{I}_{11}$,
$\mathcal{I}_{22}$ and $\mathcal{I}_{12}=\mathcal{I}_{21}$ are non-zero.
For the $x_1$ and $x_2$ components at initial and final times $t\to\pm\infty$,
when the polar angle is $\phi_0$, we have
\begin{equation}
  \lim_{t\to\pm\infty} \dot{x}_1 = {\rm v}_{\infty}\,\sin{\pm\phi_0}~,\quad
  \lim_{t\to\pm\infty} \dot{x}_2 = {\rm v}_{\infty}\,\cos{\pm\phi_0}~.
\end{equation}
Using some trigonometric identities and rearranging, one can write
the expected amplitude of the memory in the different components as
\begin{equation}\label{eq:memory_favata}
  \Delta h_{ii} = 0~,\;\text{for } i=1,2\qquad\text{and}\qquad
  \Delta h_{ij} = 8\frac{G}{c^4}\frac{E}{R}\sin{2\phi_0}~,\;\text{for } i\neq j~.
\end{equation}
One can notice that according to this crude estimate, at Newtonian order
and without spin, there is no memory effect in the plus polarization,
while there is some difference between initial and final state in
$h_{\times}$, depending on the initial settings of the binary.

\section{Gravitational waves from hyperbolic encounters}
In this section, we will present how we computed PN accurate GW templates
for both spinning and non-spinning binaries, in order to study the outcoming
memory effect and compare our result with the theoretical predictions above.

\subsection{Non-spinning binaries}

By introducing polar coordinates $(r,\phi)$ and the constant inclination
angle $\theta$ between the observer and the orbital plane, as well as
the notation $z=Gm/r$, one can show that the resulting polarization states
$h_{+,\times}$ can be written as a series of functions in terms of
$r, \dot{r}, \phi, \dot{\phi}$ as
$h_{+,\times} = -\frac{G\mu}{c^4R}\;\left(h_{+,\times}^{\text{N}}+
\frac1c h_{+,\times}^{0.5\text{PN}}+\frac{1}{c^2}h_{+,\times}^{1\text{PN}}\right)$
where the single contributions can be found in Ref.~\cite{DGGJ14}.
As an illustration, the Newtonian expressions read
\begin{subequations}\label{eq:ns_N}
\begin{align}
  h_+^N &=  2 r\dot{r}\dot{\phi} (1+C_{\theta}^2)\sin{2\phi}+(1+C_{\theta}^2)\left(z+r^2\dot{\phi}^2-\dot{r}^2\right)\cos{2\phi}+S_{\theta}^2\left(\dot{r}^2+r^2\dot{\phi}^2-z\right)~,\\
  h_{\times}^N &= 2C_{\theta}\left(z+r^2\dot{\phi}^2-\dot{r}^2\right)\sin{2\phi}-2C_{\theta}2r\dot{r} \dot{\phi} \cos{2\phi}~,
\end{align}
\end{subequations}

We now want to impose the temporal evolution of the dynamics.
Considering the 1PN hyperbolic counterpart of the Kepler equation,
and using 1PN accurate expressions for the orbital elements we
can compute $r$, $\dot{r}$, $\phi$ and $\dot{\phi}$ in terms of
the eccentricity $e_{\rm t}$, the hyperbolic mean motion $\bar{n}$,
the eccentric anomaly $v$ and the symmetric mass ratio $\eta=\mu/m$:
\begin{subequations}
  \label{eq:quasi_kepl_nonsp}
  \begin{align}
    r (v) =\; &\frac{Gm}{c^2}\frac{1}{\bar \xi^{2/3}}(e_{\rm t}\cosh v-1)\,\left\{1+\bar \xi^{2/3}\;\frac{2\eta-18-(6-7\eta)e_{\rm t}\cosh v}{6\left(e_{\rm t}\cosh v-1\right)}\right\}~, \label{1}\\
    \dot{r} (v) =\; &\bar \xi^{1/3}\frac{c\, e_{\rm t}\sinh v}{e_{\rm t}\cosh v-1}\left\{1-\bar \xi^{2/3}\frac{6-7\eta}{6}\right\}~,\\
    \phi (v) =\; &2\arctan \left[\left(\frac{e_{\phi}+1}{e_{\phi}-1}\right)^{1/2}\tanh v/2\right]\,\left\{1+\bar \xi^{2/3}\frac{3}{e_{\rm t}^2-1}\right\}~,\\ 
    \dot{\phi} (v) =\; &\frac{\bar n\, \sqrt{e_{\rm t}^2-1}}{\left(e_{\rm t}\cosh v-1\right)^2}\bigg\{1-\bar \xi^{2/3}\,\frac{\left[3-\left(4-\eta\right)e_{\rm t}^2+\left(1-\eta\right)e_{\rm t}\cosh v\right]}{\left(e_{\rm t}^2-1\right)\left(e_{\rm t}\cosh v-1\right)}\bigg\}\label{4}~,
  \end{align}
\end{subequations}
where $\bar \xi$ stands for $Gm\bar n /c^3$. Imposing eqs. (\ref{eq:quasi_kepl_nonsp})
in the expressions for the polarization states (\ref{eq:ns_N}), we obtain
the gravitational wave signals from non-spinning compact binaries during hyperbolic
encounters, at full 1PN order. In Fig.~1 we display some results for two different
values of $e_{\rm t}$.

\begin{figure}
  \begin{center}
    \includegraphics[width=\textwidth]{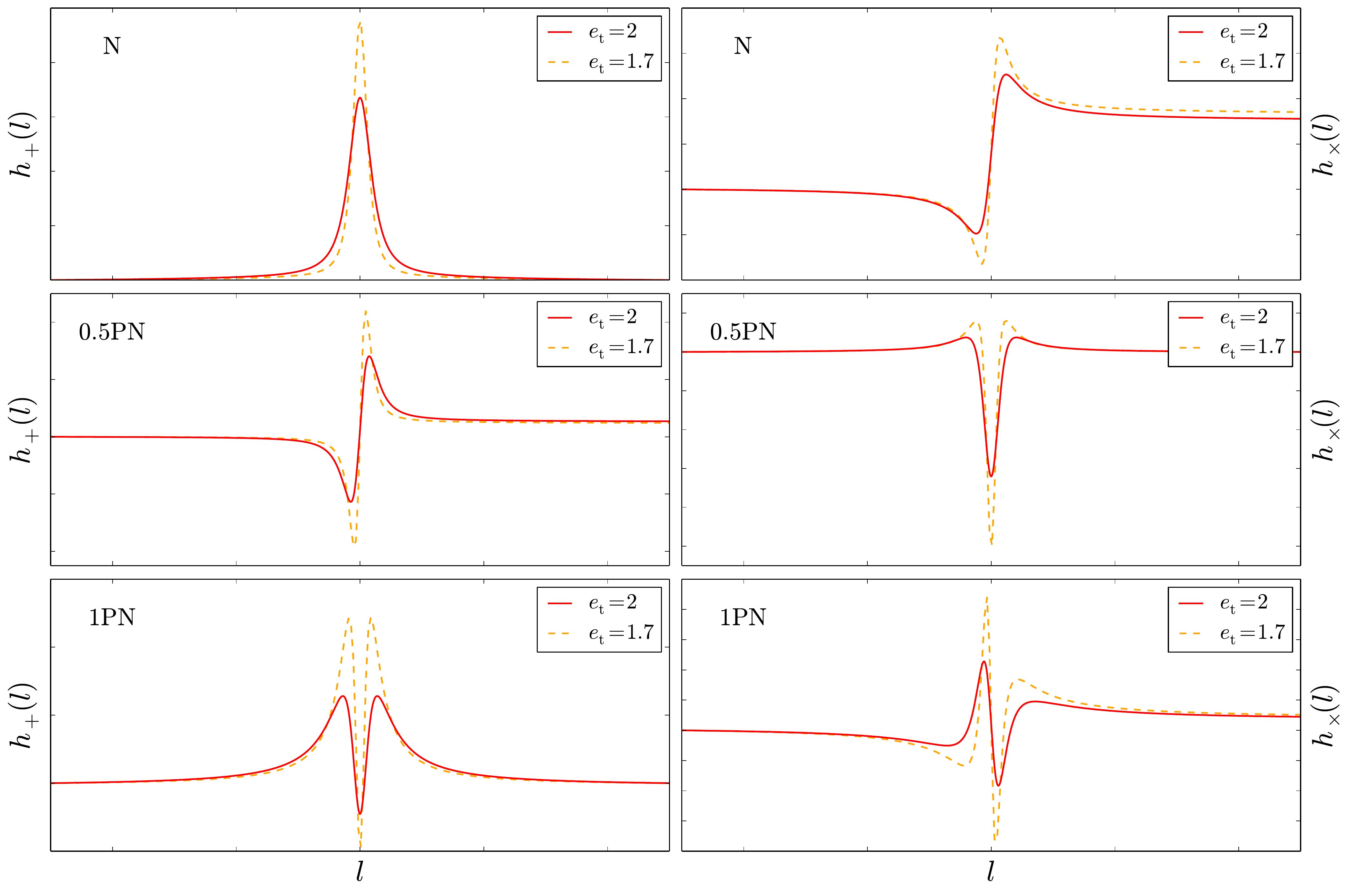}
    \label{fig:ns}
    \caption{Polarization states at Newtonian, 0.5PN and 1PN order, respectively, as functions of the
      scaled time $l$ for non-spinning compact binaries on hyperbolic orbits, in arbitrary units. The
      non-vanishing amplitude of the $\times$ polarization at late times, the so-called memory effect,
      is clearly visible at Newtonian order.}
  \end{center}
\end{figure}

A close look at Eqs.~(\ref{eq:ns_N}) allows us to extract the limit case
where $t$ (or $v$) goes to $\pm \infty$.
For simplicity, let's consider again the case where the observer
line-of-sight is perpendicular to the orbital plane, i.e. $\theta=0$.
It is not very difficult to infer that
the dynamical variables $z (v) = G\,m/r (v)$ and $\dot{\phi}(v)$, as well as the product $r(v)\dot{\phi}(v)$ go to zero.
However, the expression for $\dot{r}(v)$ does not vanish 
as $ t \rightarrow \pm \infty$, but rather tends to the finite value $\text{v}_{\infty}$.
This forces the expressions for  $h_+|_Q(t) $ and $h_{\times}|_Q(t) $ at late times to be
\begin{subequations}
\begin{align}
  h_+|_Q &= 2\frac{G\mu}{c^4R} \text{v}_{\infty}^2\, \cos{2\phi_0} \label{eq:hp_inf}~,\\
  h_{\times}|_Q  &= 2\frac{G\mu}{c^4R} \text{v}^2_{\infty}\, \sin{2\phi_0}~.\label{eq:hc_inf}
\end{align}
\end{subequations}

Note that the right hand side of Eq.~(\ref{eq:hp_inf}) is an even function of $\phi$ 
due to the presence of $\cos{2\phi }$. However, the right hand side
of Eq.~(\ref{eq:hc_inf}) is an odd function of $\phi$.
This implies that 
$\lim_{t\to +\infty} h_+|_Q(t) = \lim_{t\to -\infty} h_+|_Q(t)$, while
$\lim_{t\to +\infty} h_{\times}|_Q(t) = - \lim_{t\to -\infty}h_{\times}|_Q(t)$.

\noindent In particular, using again $\mu\,\text{v}^2_{\infty}=2E$, the amplitude difference between late and early times is
\begin{equation}\label{eq:memory_DGGJ}
  \Delta h_+ = 0~,\qquad\text{and}\qquad
  \Delta h_{\times} = 8\frac{G}{c^4}\frac{E}{R}\sin{2\phi_0}~,
\end{equation}
which is essentially the same result as in Ref.~\cite{F09} reported in eq.~(\ref{eq:memory_favata}).
This leads to the linear memory effect as visible in the $h_{\times}|_Q(l) $ plots in our Fig.~1.

\subsection{Spinning binaries}
In our recent work \cite{DGGJ14} we also included leading order spin-orbit interaction effects.
This cannot be achieved in a fully analytical way, since one obtains a set of 12 coupled
differential equations which have to be solved at the same time.
First, one needs to introduce three Eulerian angles $\Phi$, $\alpha$ and $\iota$, since
the orbital plane is precessing due to spin effects. Then we need to evolve the phasing angle
$\Phi$, the spin vectors $\bm{S}_1$ and $\bm{S}_2$ and the orbital momentum $\bm{L}$.
We were able to include radiation reaction effects, a 2.5PN order correction, by adding
two more differential equations for the evolution of the eccentricity $e_{\rm t}$ and
the mean motion $\bar{n}$. 
The system contains therefore differential equations for $\dot{\Phi}$, $\dot{\bm{S}}_1$,
$\dot{\bm{S}}_2$, $\dot{\bm{L}}$, $\dot{e}_{\rm t}$ and $\dot{\bar{n}}$. In Sec.~II of
Ref.~\cite{DGGJ14} we show these relations explicitly and explain their derivation in great detail.
\begin{figure}[h]
  \begin{center}
    \includegraphics[width=\textwidth]{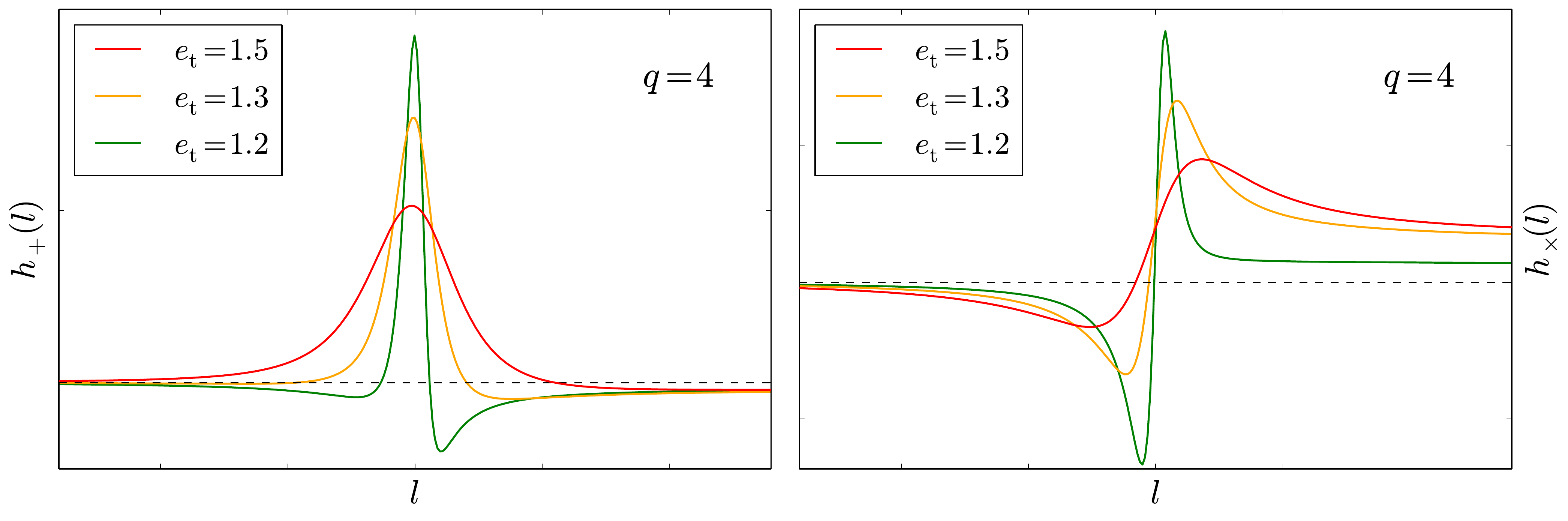}
    \label{fig:sp}
    \caption{The gravitational wave signal of spinning compact binaries on hyperbolic orbits,
      with mass ratio $q=4$ and three different values for $e_{\rm t}$. The orbital evolution
      is fully 1.5PN accurate and the influences of GW emission are taken into account. The
      linear memory effect causes the solid line waveform plots to depart from the dashed line
      after the hyperbolic passage.}
  \end{center}
\end{figure}

The key result, as one can see in Fig.~2, is that spinning binaries leave a memory imprint
on the metric in both the plus and the cross polarizations, due to the spin induced
precession of the orbital plane and the consequent asymmetry of the initial and final state
of the system.
Although small, the memory effect can be seen on the plot by comparing the initial value,
reported by the dashed line, with the final value of $h_+$ and $h_{\times}$.

\section{Conclusions}
We described a recently developed prescription to obtain ready--to--use PN-accurate GW templates
for spinning compact binaries on hyperbolic orbits, including leading order spin-orbit coupling
contributions and radiation reaction effects.
In the results we notice that for spinning binaries both the plus and the cross polarization
states exhibit a memory effect, i.e. a non-vanishing amplitude at time $t=+\infty$, whereas for
non-spinning systems only the cross polarization shows such a behaviour.
Given the interesting feature of the memory in the GW signals, we also briefly reviewed the
concept following the literature and providing a crude estimate of its amplitude at Newtonian
order, which is proven to agree with our results.

Since the waveform is strongly influenced by spin effects, the structure of such an asymmetric
burst-like signal carries many informations about the system which could be useful for parameter
estimation. It will be interesting to incorporate the 2PN order non-spinning contributions to
our 1.5PN-accurare orbital dynamics, as well as dominant order spin-spin interactions.

\end{document}